\documentclass[conference]{IEEEtran}
\IEEEoverridecommandlockouts
\usepackage{cite}
\usepackage{amsmath,amssymb,amsfonts}
\usepackage{algorithmic}
\usepackage{graphicx}
\usepackage{textcomp}
\usepackage{xcolor}
\usepackage{subcaption}
\usepackage{float}
\usepackage{stfloats} 
\def\BibTeX{{\rm B\kern-.05em{\sc i\kern-.025em b}\kern-.08em
 T\kern-.1667em\lower.7ex\hbox{E}\kern-.125emX}}

\begin{document}

\title{Scalable 28nm IC implementation of coupled oscillator network featuring tunable topology and complexity
\\

}

\author{
 S. Y. Neyaz\textsuperscript{1}, 
 A. Ashok\textsuperscript{1}, 
 M. Schiek\textsuperscript{3}, 
 C. Grewing\textsuperscript{1}, 
 A. Zambanini\textsuperscript{1}, 
 S. van Waasen\textsuperscript{1,2}\\
 \textsuperscript{1}Central Institute of Engineering, Electronics and Analytics – Electronic Systems (ZEA-2)\\
Forschungszentrum Jülich, 52425 Jülich, Germany\\
 \textsuperscript{2}Faculty of Engineering, Communication Systems, University of Duisburg-Essen, 47057 Duisburg, Germany\\
 \textsuperscript{3}Peter Grünberg Institut (PGI-14)
 Forschungszentrum Jülich, 52425 Jülich, Germany
 
}

\maketitle

\begin{abstract}
Integrated circuit implementations of coupled oscillator networks have recently gained increased attention. The focus is usually on using these networks for analogue computing, for example for solving computational optimization tasks. For use within analog computing, these networks are run close to critical dynamics. On the other hand, such networks are also used as an analogy of transport networks such as electrical power grids to answer the question of how exactly such critical dynamic states can be avoided. However, simulating large network of coupled oscillators is computationally intensive, with specific regards to electronic ones. We have developed an integrated circuit using integrated Phase-Locked Loop (PLL) with modifications, that allows to flexibly vary the topology as well as a complexity parameter of the network during operation. The proposed architecture, inspired by the brain, employs a clustered architecture, with each cluster containing 7 PLLs featuring programmable coupling mechanisms. Additionally, the inclusion of a RISC-V processor enables future algorithmic implementations. Thus, we provide a practical alternative for large-scale network simulations both in the field of analog computing and transport network stability research.

\end{abstract}


\section{Introduction}

Coupled oscillators with adaptive properties mimic many real-world networks including power grids where generators and consumers are represented by oscillators~[1], and neuroscience, where such networks describe the brain rhythm dynamics~[2]. Various mathematical models, such as Winfree, Kuramoto, Sakaguchi-Kuramoto, Stuart-Landau, FitzHugh-Nagumo, have been associated with these networks and extensively studied in the field of complex non-linear dynamics~[3], with a focus on coupling strength, delay, and non-linearity, between the oscillators. In an experimental study [4], the impact of coupling strength, and phase lag in a coupled metronome system (3 metronomes) was studied. The analysis identified phase lag in the coupling as a key complexity parameter for the system. As the phase lag increased towards \(\frac{\pi}{2}\), chimera states emerged, characterized by the coexistence of synchronized and desynchronized regions. In the context of IC, the network synchronization dynamics is independent of the type of oscillator but depends on the type of coupling and delay [5].
\\
Another aspect of coupled oscillators has been utilized to develop applications that exploit their synchronization behavior, one of them being Oscillatory Neural Networks (ONNs). These networks harness the synchronization of oscillators for specific tasks, such as pattern matching [6]. In such networks, the non-boolean logic approach is used where the phase relationship among oscillators defines the system's state. One of the key advantages of using non-boolean logic lies in minimal operating voltage levels that enable lower power consumption [5].
 \\
 ONNs have been successfully applied to various fields, including edge detection~[7], edge computing~[8], to optimization problems~[9]. In~[10], a proposed solution for the graph coloring problem utilizes a continuous-time dynamical system based on coupled oscillators to assign colors to graph vertices. The synchronization behavior of the oscillators determines the colors. However, challenges arise as the system's complexity grows with the number of nodes (oscillators), leading to extended simulation times. Furthermore, a proposal to use relaxation oscillators incorporating inductors will result in bulky hardware implementations. To address this, more compact approaches using vanadium dioxide ($VO_{2}$) crossbar oscillators have been implemented~[11].
 Despite these advances, scalability remains a limitation, as this implementation is restricted to nine oscillators. To overcome this, the authors propose graph partitioning to create more manageable systems. 
In addition, smaller-scale experiments using off-the-shelf components like diodes and ring oscillators have been used to study dynamical behaviors such as bifurcation and chaos [12]. These studies hold potential for applications in fields like random number generation and cryptography.
 \\
In [13], a Boolean phase oscillator based on a modified ring oscillator is implemented on an FPGA. This system closely resembles the Kuramoto model and provides a digital framework for studying various nonlinear dynamical regimes.
Our goal is to develop an integrated circuit solution that enhances scalable, practical studies of complex dynamical systems. The proposed coupled oscillator system, featuring a RISC-V processor aims at supporting nonlinear dynamics research, including bifurcation studies and algorithm development. With a scalable design of clustering approach and tunable delay $\tau$ (this is the complexity parameter), this system provides a versatile platform for exploring various dynamical behaviors beyond the constraints of large-scale simulations.
 By enabling real-time operation, it overcomes the long simulation times noted in recent research [10]. Additionally, the hybrid mode allows ‘‘on-the-fly" parameter adjustments, offering flexible, and real-time adaptability.

 \begin{figure}[t]
 \centering
 \includegraphics[width=\linewidth]{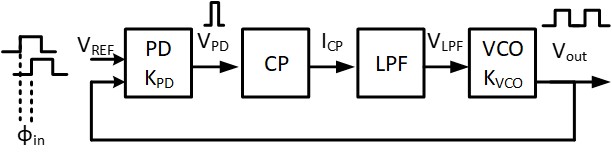} 
 \caption{Block diagram of the PLL, where PD is the phase detector with a gain of $K_{PD}$, CP is the charge pump with gain $I_{CP}$, LPF is the loop filter and VCO with a gain $K_{VCO}$ is voltage controlled oscillator. }
 \label{fig:fig_1}
 \vspace{-0.3cm}
\end{figure}

\vspace{-0.1cm}

\section{Coupled Oscillator using PLL}

\subsection{Methodology}

The Kuramoto model is a well-known mathematical model describing the behavior of coupled oscillators, and is known to exhibit rich dynamics with the inclusion of a delay parameter [14]. Time delays are crucial for accurately mimicking real-world scenarios, making their inclusion essential for capturing realistic interactions. Therefore, the extended model in (1), which incorporates delay, provides deeper insights into these complex dynamics.

\begin{equation}
\frac{d\theta_i}{dt} = \omega_i + \frac{K}{N} \sum_{j=1}^{N} \sin(\theta_j(t - \tau) - \theta_i(t))
\end{equation}
\begin{itemize}
 \item \(\theta_i(t)\): Phase of the \(i\)-th oscillator at time \(t\).
 \item \(\omega_i\): Natural frequency of the \(i\)-th oscillator.
 \item \(K\): Coupling strength between oscillators.
 \item \(N\): Total number of oscillators.
 \item \(\tau\): Time delay in the interaction between oscillators.
\end{itemize}

As can be seen from (1), the information processing is inherently phase-based, where the instantaneous frequency of the oscillator \(\omega_i\) is influenced by the phase relationships among the oscillators. Similarly, it is well-known that the instantaneous frequency \(\omega_{out}\) of a Voltage Controlled Oscillator (VCO) is affected by the control voltage $V_{ctrl}$ as represented by (2).
\begin{equation}
\omega_{out}= \omega_{0} +K_{VCO}V_{ctrl}
\end{equation}
\begin{itemize}
 \item \(\omega_{out}\): Output frequency of the VCO (rad/s).
 \item \(\omega_{0}\): Natural frequency of the VCO (rad/s).
 \item \(K_{VCO}\): VCO sensitivity (Hz per volt).
 \item \(V_{ctrl}\): Control voltage of the VCO (V).
\end{itemize}
In this context, our approach is to make $V_{ctrl}$ dependent on phase differences among oscillators, thereby directly linking phase information to the frequency control mechanism.
\\
\begin{figure}[h]
 \centering
 \includegraphics[width=0.8\linewidth]{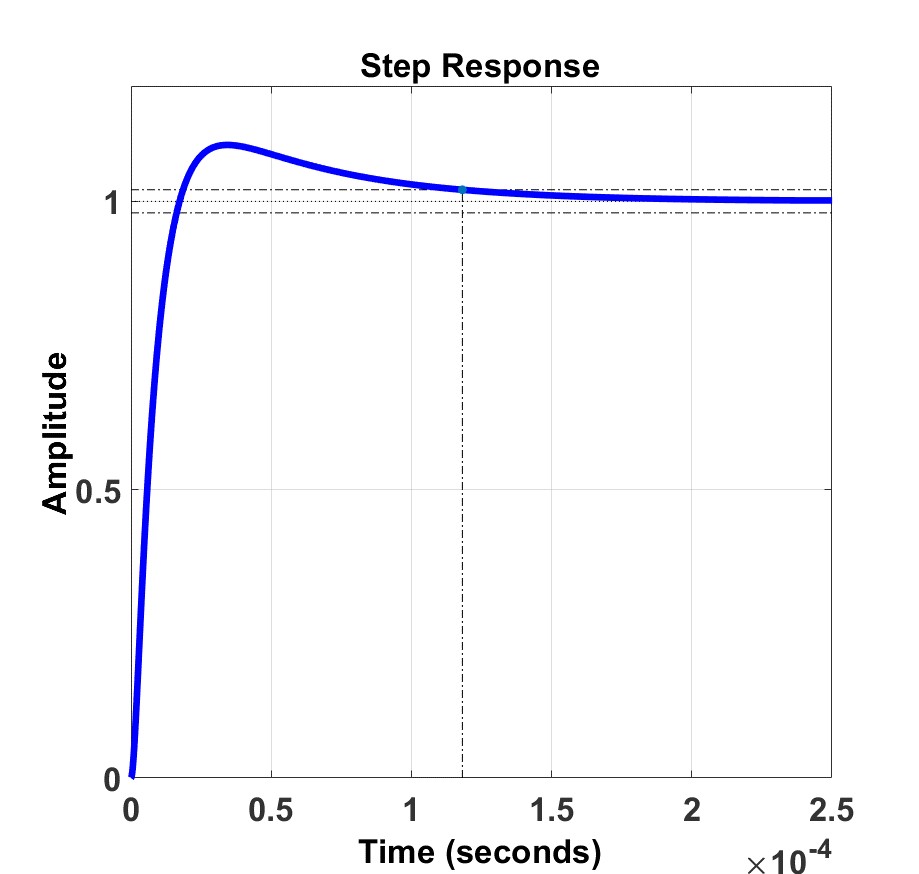} 
 \caption{Step response of type-2 3\textsuperscript{rd} order PLL}
 \label{fig:fig_2}
 \vspace{-0.3cm}  
\end{figure}
\\
Equation (1) highlights the phase comparison which is not explicitly present in (2), and therefore modifications are necessary to ensure its alignment with (1). A phase comparison can be performed via an XOR gate, with an output proportional to the phase difference between the oscillators. This signal can influence the $V_{ctrl}$, after appropriate filtering, leading to a type-1 PLL [15].
\\
 An important characteristic of a PLL is its acquisition range which is dependent on the loop bandwidth [15]. A narrow loop bandwidth is necessary to prevent any spurious tones from influencing the VCO. However, this also limits the PLL to capture frequencies within a very narrow range, which may not meet the requirements of (1). To increase the bound of the acquisition range, a type-2 PLL of order 3 can be used, whose transfer function and block diagram are shown in (3) and Fig. 1 respectively.
\begin{figure}[b]
 \centering
 \includegraphics[width=0.75\linewidth]{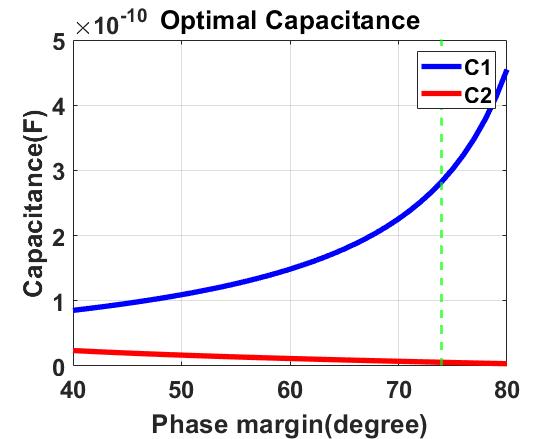} 
 \caption{Optimal capacitance for a certain Phase margin $\phi_{m}$ }
 \label{fig:fig_3}
 \vspace{-0.7cm}
\end{figure}

\begin{equation}
\frac{\phi_{\text{out}}}{\phi_{\text{in}}} =
\frac{K_{PD} \cdot K_{VCO} \cdot \left( \frac{s + \omega_z}{C_2} \right)}{s^3 + s^2 \omega_p + \frac{K_{PD} \cdot K_{VCO} \cdot s}{C_2} + \frac{K_{PD} \cdot K_{VCO} \cdot \omega_z}{C_2}}
\label{eq:transfer_function}
\end{equation}

\
\begin{itemize}
 \item \(\phi_{\text{in}}; \phi_{\text{out}}\): Input and Output phase, representing the phase of the input and output signal,

 \item \(\omega_z; \omega_p\): Zero and Pole frequency 

\end{itemize}

\subsection{PLL based Coupled Oscillator Specifications}
As observed from (1), the coupled oscillator system works on the principle that each oscillator is influenced by the superposition of phase differences relative to the other oscillators. In this system, the inputs to each oscillator are the phase differences between itself and the others, which collectively determine its behavior. Therefore, modifications to the conventional type-2 PLL need to be made for the multi-input phase comparison. A divider is also absent in the feedback loop (N=1) as the input and output frequencies are the same. These modifications and their implications will be elaborated upon in subsequent sections.
\\
Biological neural systems are characterized by low-frequency signal processing [16]. Based on this principle, the VCO of PLL has been designed to operate at around 10 MHz, adapted for electronic systems requiring higher processing speeds while maintaining a biologically-inspired coupling mechanism.


Another key aspect of the network is ensuring the robustness of the PLL, which translates to stability. This requirement is addressed by designing the PLL with a phase margin $\phi_{m}$= 74$^\circ$ resulting in an overdamped system. However, this large phase margin leads to an increased settling time of approximately 120 µs, as shown in Fig. 2. This trade-off between stability and settling time is intentionally leveraged in the proposed coupling where two schemes, CS-1 and CS-2, which are described in detail in later sections (refer to section III).
\\
For such an overdamped system, filter bandwidth is selected less than 1\% of the reference frequency. This also suppresses high-frequency components and noise, preventing any undesired coupling. Since the power consumption scales with the number of oscillators, charge pump current $I_{CP}$ is kept low. Additionally, $K_{VCO}$ has been kept low (see Table 1) to reduce power supply-induced coupling and ensure that CS-1 or CS-2 remains the primary source of coupling.
\\
The design of the PLL for coupled oscillators is implemented in 28nm bulk CMOS technology and the PLL blocks are designed with careful consideration of the limited active area of the chip. 
An optimal capacitance is selected for the required phase margin as shown in Fig. 3.
\begin{figure*}[b]
 \centering
 \includegraphics[width=0.8\textwidth]{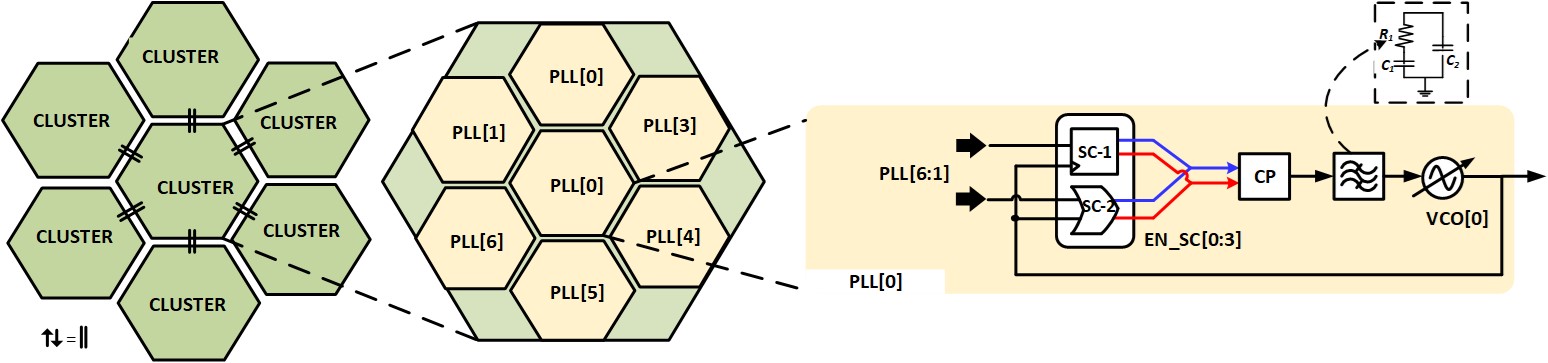}
 \caption{Cluster architecture of coupled oscillator with 7 clusters, and 7 PLLs in each cluster. Two coupling schemes CS-1 and CS-2 as well as a programmable delay}
 \label{fig:wide-image1}
\end{figure*}

\begin{table}[h!]
\vspace{0.1cm}
\centering
\caption{Design Parameters and Values}
\begin{tabular}{l l}
\hline
\textbf{Parameters}  & \textbf{Values}  \\ \hline
Phase margin (\( \phi_m \)) & 74°     \\
Loop Bandwidth (\( \omega_{BW} \)) & \( 2\pi \times 27 \times 10^3 \, \text{rad/s} \)

  \\
Kvco      & 3 MHz/V    \\
Divider ratio (N)   & 1     \\
I$_{CP}$     & 1.34 µA    \\ \hline
\vspace{-0.6cm}
\end{tabular}
\end{table}

\section{System architecture}

As the number of oscillators increase, the interconnection density increases, posing a challenge to the scalability of these systems. In our approach, as illustrated in Fig. 4, we address this issue by following a clustered architecture where the oscillators are split into clusters. The intra cluster coupling allows for either an all-to-all coupling or sparse connectivity depending on our requirements, while inter cluster coupling can be nearest neighbor coupled.




The programmability of the proposed system offers flexibility for exploring various operational scenarios, such as assessing the impact of adjusting or removing PLLs from a cluster. Such programmability enhances the system's adaptability for various use cases, including real-time adjustments and fault-tolerant designs in both oscillatory neural networks and critical infrastructure systems like power grids, communication networks, and beyond [17]. This flexibility can drive innovations in how interconnected systems are managed and optimized.

\vspace{-0.4cm}
\subsection{Coupling schemes}
The PLL within a cluster can be coupled through two mechanisms as illustrated in Fig. 5. In coupling Scheme 1 or CS-1, as depicted in Fig. 5(a), the outputs $V_i$ from all the individual PLLs within a cluster are compared with the $V_k$ of PLL$_k$, where $V_k$ is the output voltage associated with VCO of PLL$_k$. The resulting pulses are then processed through multi-input logic gates allowing superimposed output.
The implemented logic dictates that the cumulative effect of all leading PLLs in a time interval accelerates the PLL P$_k$ through an AND gate. In contrast, the outputs from lagging PLLs, routed through an OR gate, contribute to slowing down the same PLL P$_k$. A symbolic representation of CS-1 is also shown in Fig. 5(a).
\begin{figure*}[ht]
 \centering
 \includegraphics[width=0.8\textwidth]{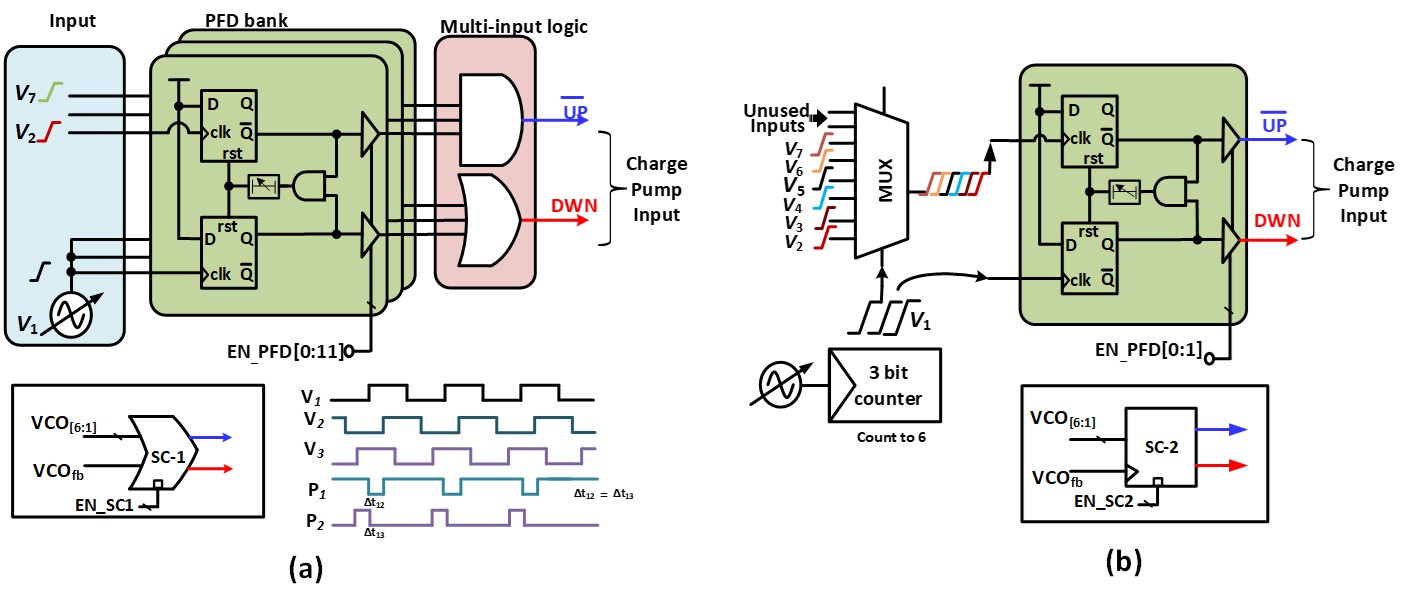}
 \caption{(a) Coupling scheme 1 or CS-1 (b) Coupling scheme 2 or CS-2}
 \label{fig:wide-image2}
\end{figure*}
\\
For a scenario where PLL$_2$ and PLL$_3$ exhibit phase differences, leading and lagging PLL$_1$ respectively by equal amounts. In this case, PLL$_2$ accelerates PLL$_1$ through the AND gate, while the influence of PLL$_3$ decelerates PLL$_1$ via the OR gate. Since the phase deviations of the PLLs concerning PLL$_1$ are symmetrical but opposite in effect, the contributions cancel out each other, and there is no net phase shift. Quantitatively, in a small time interval \(\Delta t_a\)
as 
\begin{equation}
 \Delta t_{12} = -\Delta t_{13} 
 \end{equation}
 
\begin{equation}
\Delta \phi_{1} = 2\pi \times \frac{\left( \Delta t_{12} + \Delta t_{13} \right)}{T}
\end{equation}
where $\Delta t_{12}$ and $\Delta t_{13}$ are time delays of PLL$_2$ and PLL$_3$ with respect to PLL$_1$.
Therefore, in a small time interval \(\Delta t_a\), there is effectively no change in the instantaneous phase \(\Delta \phi _{1}\) of PLL 1.
CS-1 shares important similarities with the dynamics of Kuramoto oscillators. Oscillators with equal leading and lagging phase differences result in no net phase shift for the central oscillator.
\\
Coupling scheme 2 or CS-2 is designed around the principle that the designed PLL has a longer settling time (\(t_s\)), shown in Fig. 5(b).
Within a cluster, phase comparison of each PLL with others is carried out serially using a multiplexer. This multiplexer is driven by a 3-bit counter which is triggered by the VCO edge as shown in Fig. 5(b).
\subsection{VCO}
The low $K_{VCO}$ is achieved in the design using a feedback structure as shown in Fig. 6. The current through the resistor $I_{ctrl}$ is proportional to the output of the loop filter $V_{ctrl}$. $I_{ctrl}$ is mirrored to the VCO core which is a standard CMOS ring oscillator of back-to-back inverters. The programmability 
\begin{figure}[htbp]
 \centering
 \includegraphics[width=0.95\linewidth]{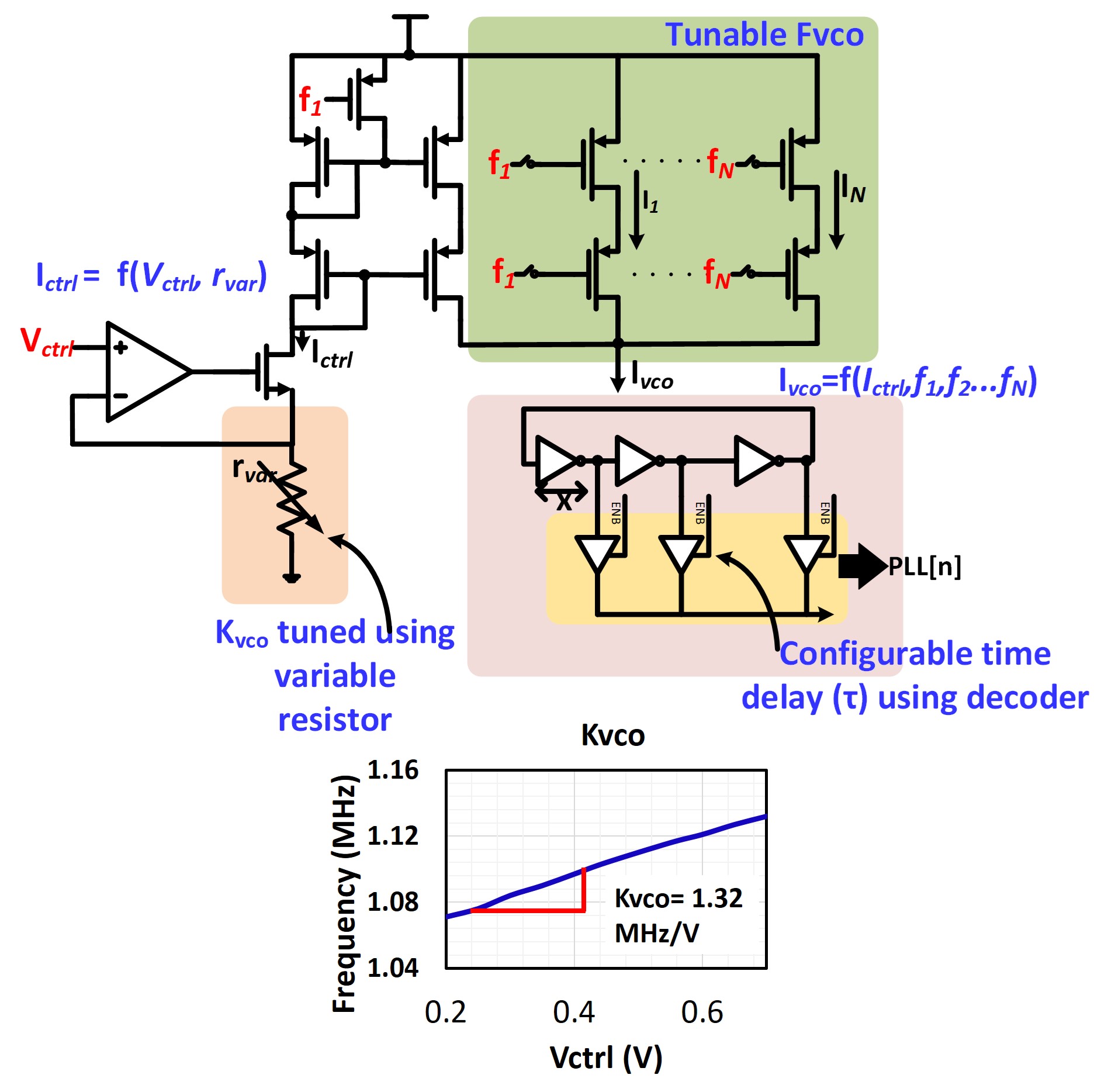} 
 \caption{Tunable VCO with Time Delay Mechanism for PLL and $K_{VCO}$}
 \label{fig:example}
 \vspace{-0.2cm} 
\end{figure}
of the coupled oscillators comes from the variation of resistor $r_{var}$ as well as mirroring of $I_{ctrl}$ leading to $K_{VCO}$ and central frequency modifications. Moreover, a delay $\tau $  in the coupling is implemented using tristate buffers at each stage. This allows for dynamic adjustment of signal propagation delay.

\begin{figure}[h]
 \centering
 \includegraphics[width=0.8\linewidth]{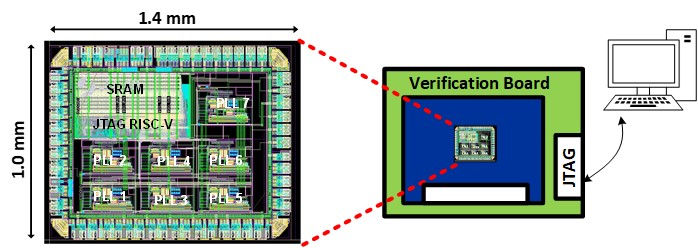} 
 \caption{Layout of 28nm coupled oscillator in 28nm CMOS with JTAG interface}
 \label{aaa}
 \vspace{-0.5cm}
\end{figure}

\section{Conclusion \& Outlook }
The proposed PLL-based coupled oscillator at 28nm process technology offers capabilities similar to that of the Kuramoto model with delay, which emulates many real-world systems. 
 Additionally, the coupling schemes in this design offer flexibility in selecting and comparing different types of coupling, which aids in determining critical parameters, without long-time simulations. The chip layout, as shown in Fig. 7, includes a JTAG interface and a RISC-V processor, laying the groundwork for future algorithm implementation. The tapeout of the chip was in July 2024 and after system verification the first field of application of the chip will be for network stability analysis of power grids.




\vspace{12pt}


\begin{thebibliography}{00}
\bibitem{b1} T. Coletta, and P. Jacquod, “Linear stability and the Braess paradox in coupled-oscillator networks and electric power grids,'' Phys. Rev. E 93, 032222, March 2016
\bibitem{b2} C. Bick, M. Goodfellow, C.R. Laing, and E. A. Martens, “Understanding the dynamics of biological and neural oscillator networks through exact mean-field reductions: a review,'' J. Math. Neurosc. 10, 9 (2020).
\bibitem{b3} P. Ashwin, S. Coombes and R. Nicks, “Mathematical frameworks for oscillatory network dynamics in neuroscience. J. Math,'' Neurosc. 6, article number: 2, 2016
\bibitem{b4} P. Ebrahimzadeh, M. Schiek, P. Jaros, T. Kapitaniak, S.van Waasen, Y.Maistrenko, ``Minimal chimera states in phase-lag coupled mechanical oscillators.,'' The European Physical Journal Special Topics, Japan, vol. 229, p.2205–2214, 2020,
\bibitem{b5} Y. Fang, C. N. Gnegy, T. Shibata, D. Dash, D. M. Chiarulli, and S. P. Levitan, ‘‘Non-Boolean associative processing: Circuits, system architecture, and algorithms,’’ EEE Journal on Exploratory Solid-State Computational Devices and Circuits. 1. 1-1.
\bibitem{b6} D. E. Niknonov et al., ``Coupled oscillator associative memory operation for pattern recognition,'' IEEE Journal on Exploratory Solid State Computational Devices and Circuits 1, 85, 2015.
\bibitem{b7}M. Abernot, A. Todri-Sanial, ``Simulation and implementation of two-layer oscillatory neural networks for image edge detection: bidirectional and feedforward architectures,'' Neuromorph. Comput. Eng. 3:014006, 2023.
\bibitem{b8} M. Abernot, A. Todri-Sanial, ``Oscillatory neural networks implemented on FPGA for edge computing applications,'' 26th Design, Automation and Test in Europe Conference, Apr 2023.

\bibitem{b9} T. Wang, L. Wu, P. T. Nobel, and J. Roychowdhury, ``Solving combinatorial optimization problems using oscillator based Ising machines,'' Natural Computing. vol.20, pp. 1-20,2021. 
\bibitem{b10} W. W. Chai, “Graph coloring via synchronization of coupled oscillators,'' IEEE Trans. Circuits Syst. I45(9), p.974–978, 1998.
\bibitem{b11} O. Maher, et al. ``A CMOS-compatible oscillation-based VO2 Ising machine solver,'' Nat Commun 15, 3334, 2024.
\bibitem{b12} H. Mondal, A. Pathak, T. Banerjee, M. Mandal, ``Ring oscillators under nonlinear coupling: Bifurcation and Chaos,'' ECTI Transactions on Electrical Engineering, Electronics, and Communications, 2023. 
\bibitem{b13} D. P. Rosin, ``Dynamics of complex autonomous boolean networks,'' Springer Theses, 2015
\bibitem{b14} M. K. S. Yeung and S. H. Strogatz, ``Time delay in the Kuramoto model of coupled oscillators,'' Proceedings of the Physical Review Letters Conference, vol. 82 (3), p. 648-651, 1999.
\bibitem{b15} B. Razavi, Design of CMOS Phase-Locked Loops: From circuit level to architecture level. McGraw-Hill, 2002.
\bibitem{b16} A . K. Engel, P. Fries P, W. Singer, ``Dynamic predictions: oscillations and synchrony in top-down processing,'' Nat Rev Neurosci., vol. 2(10), p.704-16, 2001
\bibitem{b17} F. Dörfler, and F.Bullo, ``Synchronization in complex networks of phase oscillators: A survey, ``Automatica, vol. 50 (6), p. 1539-1564, 2014.
\bibitem{b18} Hoppensteadt, F. C. \& Izhikevich, E. M. ``Oscillatory neurocomputers with dynamic connectivity, '' Physical Review Letters 82, 2983, 1999.
\end{thebibliography}
\end{document}